\documentclass[12pt]{article}
\usepackage{amsmath}
\usepackage{graphicx,psfrag,epsf}
\usepackage{enumerate}
\usepackage{natbib}
\usepackage[hyphens]{url}
\usepackage{hyperref}
\usepackage[textsize=tiny]{todonotes}
\usepackage{float}
\usepackage{xcolor,listings,realboxes,fancyvrb} 
\definecolor{mygray}{rgb}{0.8,0.8,0.8}
\usepackage[doublespacing]{setspace} 

\newcommand{\blind}{1}

\def\spacingset#1{\renewcommand{\baselinestretch}%
{#1}\small\normalsize} \spacingset{1}
\begin{document}



\if1\blind
{
 \title{Using GitHub Classroom To Teach Statistics}
 \author{Jacob Fiksel\\
  Department of Biostatistics, Johns Hopkins Bloomberg School of Public Health\\
  and \\
  Johannna S. Hardin \\
  Department of Mathematics, Pomona College \\
  and\\
  Leah R. Jager \\
  Department of Biostatistics, Johns Hopkins Bloomberg School of Public Health\\
  and\\
  Margaret A. Taub \\
  Department of Biostatistics, Johns Hopkins Bloomberg School of Public Health}
 \maketitle
} \fi

\if0\blind
{
 \bigskip
 \bigskip
 \bigskip
 \begin{center}
   {\LARGE\bf Using GitHub Classroom To Teach Statistics}
\end{center}
 \medskip
} \fi

\bigskip
\begin{abstract}
Git and GitHub are common tools for keeping track of multiple versions of data analytic content, which allow for more than one person to simultaneously work on a project. GitHub Classroom aims to provide a way for students to work on and submit their assignments via Git and GitHub, giving teachers an opportunity to teach these version control tools as part of their course. In the Fall 2017 semester, we implemented GitHub Classroom in two educational settings--an introductory computational statistics lab  and a more advanced computational statistics course. We found many educational benefits of implementing GitHub Classroom, such as easily providing coding feedback during assignments and making students more confident in their ability to collaborate and use version control tools for future data science work. To encourage and ease the transition into using GitHub Classroom, we provide free and publicly available resources--both for students to begin using Git/GitHub and for teachers to use GitHub Classroom for their own courses.

\end{abstract}

\noindent%

\vfill

\newpage
\spacingset{1.45} 
\section{INTRODUCTION AND MOTIVATION}
\label{sec:intro}

As more businesses, governments, and researchers make analysis-driven decisions, students of statistics and data analysis should be taught how to collaborate with others in managing data, code, and results that are part of a reproducible analysis pipeline. Based on an industry-wide survey of over 16,000 data scientists conducted by Kaggle, 58.4\% of the 16,000 respondents said that Git was the main tool used for sharing code in their workplace \citep{kaggle}. Using a reproducible workflow in statistics is vital to a complete data analysis, yet for faculty and students with limited computing background, learning version control tools such as Git can be difficult and intimidating. We outline some of the ways that the Git workflow can be implemented in statistics courses at all levels, and we believe that our work and that of others will help ease the larger statistics community into using GitHub across the entire statistics curriculum.

Recent work presents clear motivation and examples of using Git for statistics and data analysis. Jenny Bryan has created an impressive and comprehensive website for using GitHub with RStudio \url{http://happygitwithr.com}. Furthermore, she argues that incorporating Git and GitHub into data science workflows is considered best practice, and she provides thoughtful advice on how to conceptualize the GitHub workflow \citep{bryan2018}. Other work \citep{rundel2018} describes a method of integrating Git and GitHub into statistics courses targeted towards students with computational backgrounds. However, the previous literature does not describe best practices for handling the potential multitudes of classroom assignments nor for introducing version control tools in statistics courses with non-mathematical emphases, such as those in public health or the life sciences.

We believe that version control can and should be integrated into all statistics courses, no matter the target audience. In order to achieve this goal, we advocate for the use of GitHub Classroom. GitHub Classroom is software that aims to provide a way for students to work on and submit their assignments via Git and GitHub, while also giving teachers an opportunity to present version control tools as part of the course material. Benefits of using GitHub Classroom in an educational setting include unlimited private repositories for student work, in compliance with United States FERPA (Family Educational Rights and Privacy Act of 1974) regulations, flexible workflows for grading assignments, and ease of distribution of starter materials for various assignments. 

However, there remains a reasonably steep learning curve associated with GitHub Classroom, and implementing it can introduce logistical challenges with respect to weekly homework assignments and projects. As it is a new tool, there are no well-documented and simple workflows published that outline how to best use GitHub Classroom. To that end, we have created easy-to-use and publicly available resources that give step-by-step instructions on implementing GitHub Classroom in any statistics or data analysis course. Our instructions not only help instructors  set up their own GitHub Classroom, but also help students learn how to use Git and GitHub. This removes the need for instructors to develop their own lesson plans from scratch on how to teach Git and GitHub.

The main goal of this article is to expand on existing GitHub resources in order to share our recommended workflow for using GitHub Classroom as an educational tool and class management system. We begin in Section \ref{sec:experiences} with describing our experience in implementing GitHub Classroom in two educational settings -- an introductory computational statistics (ICS) lab  and a more advanced computational statistics (ACS) course. In Section \ref{sec:guides}, we describe the open source and publicly available tools and guides we have developed for using GitHub Classroom. We have teacher-focused resources, which are targeted towards instructors (of all subjects) who wish to set up a GitHub Classroom, and student-focused resources, which can be distributed by instructors. Both resources provide visual guides to Git, GitHub, and GitHub Classroom for instructors and students who have never used version control before. The remainder of Section \ref{sec:resources} discusses key aspects of our workflow for using GitHub Classroom, which are supplemented by our guides. We conclude the article with a brief summary and discussion. 

\section{EXPERIENCES WITH GITHUB CLASSROOM}
\label{sec:experiences}

Before delving into the specifics of our GitHub Classroom workflow, we first want to show that the tool has educational value, both pedagogically for instructors and practically for students. We do this by describing our experiences with using GitHub Classroom in two statistics education settings, one introductory and one advanced. Based on our experiences, we offer practical suggestions for introducing and motivating GitHub Classroom to students, which are targeted to student background knowledge. These more general suggestions supplement the more detailed GitHub Classroom guides that we provide later in the article. 

\subsection{Student Background in Our Two Statistics Courses}
\label{sec:stntbkgrd}

Around 20 students took the ICS course which was an optional one-credit course associated with an introductory course in biostatistics aimed at public health majors. This course met one hour per week. The vast majority of the students in this course had never done any coding, had never used a command-line interface, and had never heard of Git or GitHub. Seventy students took the ACS course, which met three hours per week. While most students in the ACS course had used R and RStudio before, very few had any knowledge or experience with Git or GitHub. In both classes, all assignments were completed using the R statistical computing language and coding was done inside the RStudio IDE.

The two classes had very different curricula, due to the vast differences in the statistical and programming knowledge bases of the two student groups. For example, the first assignment in the ICS lab had the students write a for loop in which they simulated from a normal distribution and stored the means of the simulated data sets in a vector. On the other hand, the first assignment in the ACS course was an inferential analysis of a real dataset. We believe the two classes span a wide spectrum of the typical undergraduate material taught by many statistics departments.

\subsection{Instructor Background in Our Two Statistics Courses} \label{sec:instrbkgrd}

Keeping in mind that we are training the next generation of statisticians, we are well aware of (and most of us belong to!) the previous generation of statisticians. That is, some statistics instructors may want to teach Git/GitHub to their students but not be comfortable using it in their own work. None of us were comfortable with the complete suite of Git/GitHub functionality, and all of us learned quite a bit when implementing GitHub Classroom in our courses. Our combined experience is that a rudimentary understanding of commit, push, and pull, along with a TA who has worked independently with Git, is sufficient for a successful introduction of Git/GitHub into a statistics class at any level. We refer instructors to familiarize themselves with Git and GitHub to the supplementary material in \citet{bryan2018}.

\subsection{Pedagogical and Practical Benefits of GitHub Classroom}
\label{sec:benefits}

Because there are start-up costs associated with introducing GitHub and GitHub Classroom to students, it is worth discussing why instructors should implement GitHub Classroom to run a course, rather than using the default University course management system. An immediate advantage is for classes that have group projects. With GitHub Classroom, instructors can easily assign groups of students to teams and give each team their own GitHub repository within a GitHub Classroom. Students can then use Git and GitHub to collaborate on a project, just as they would in an academic or industry research project. Because teachers can see each student's commit history, it is easy to see how each student contributed to the project. The ACS course took full advantage of group work on GitHub and had students complete team projects for their final evaluation. 

\begin{figure}[H]
    \centering
    \includegraphics[page=1, width=1.1\linewidth]{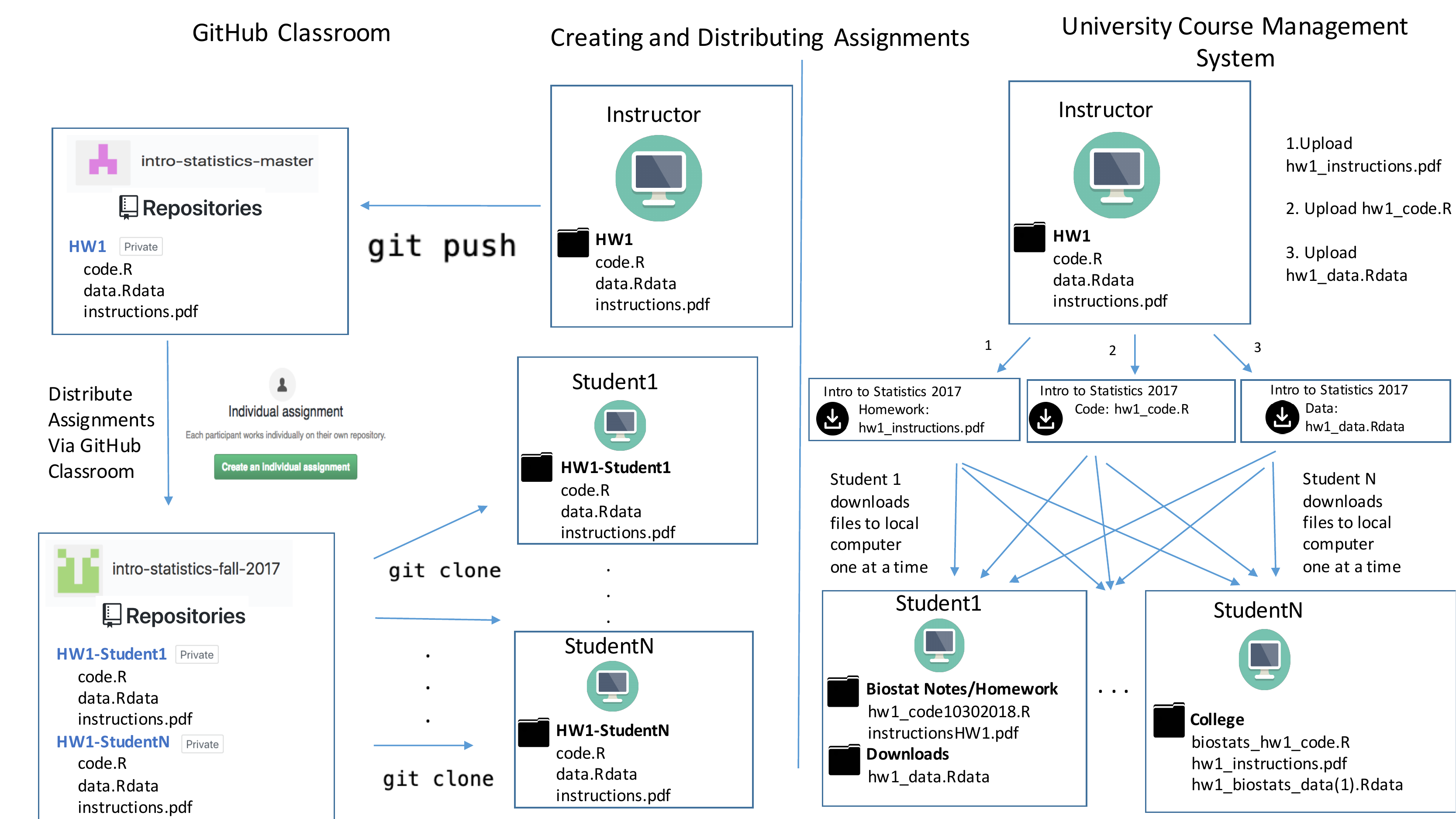}
    \caption{\textbf{Creation and distribution of assignments with GitHub Classroom versus a University CMS}. In both settings, the instructor begins with the homework assignment, which contains (starter) code, data, and instructions, on their local computer. With GitHub Classroom, the instructor pushes all parts of the assignment to the GitHub master organization. Using the GitHub Classroom interface, the instructor can create a homework repository for each student with a click of a button. Students then use {\ttfamily git clone} to download the homework assignment onto their local computers, maintaining the same directory structure and file names. Instructors using a CMS would have to upload each piece of the assignment individually. Each piece of the assignment is then downloaded individually by students. Because the students do not clone the whole assignment directory into their local computers, students can end up with different directory structures and/or different file names, which can result in difficulty running starter code and producing reproducible analyses}
    \label{fig:gh_create}
\end{figure}

Even without group projects, however, GitHub Classroom has benefits over standard academic course management systems (CMS) such as Blackboard, Sakai, and CoursePlus \citep{zagalsky2015emergence}. First, it reduces the amount of work and chances for errors during the assignment creation workflow (see Section \ref{sec:assignments}), relative to a standard CMS. We diagram how GitHub Classroom simplifies the assignment creation process in Figure \ref{fig:gh_create}. Instructors can maintain starter material for assignments on their local computer and can give students their own versions of the starter material with the push of a single button, as opposed to individually uploading each piece of the assignment to the CMS for students to download. Students can use Git to bring the full assignment onto their personal computer, without individually downloading each part of the assignment. We see in Figure \ref{fig:gh_create} how individually downloading files can result in different file structures on each students' computer, which makes it far more likely that the resulting code will have errors or not be reproducible. 

GitHub Classroom can easily be used to provide feedback on code while students are working on individual assignments. In previous iterations of the ICS lab, students asked for coding help by either emailing code and data as an attachment, or by scheduling an in-person appointment. To provide help remotely, the professor would have to 

\begin{enumerate}
\item Download the code and data
\item Ensure that the data is in the correct directory, as specified by the code
\item Run the code and provide feedback
\item Email all documents back as attachments
\end{enumerate}
	By using GitHub Classroom, we provide all students with the same initial directory for each assignment. Instructors are automatically added as collaborators to each repository, and can provide feedback through GitHub's push and pull functionality. We describe the exact workflow later in Section \ref{sec:feedback}.
	
\begin{figure}[H]
    \centering
    \includegraphics[page=2, width=1.1\linewidth]{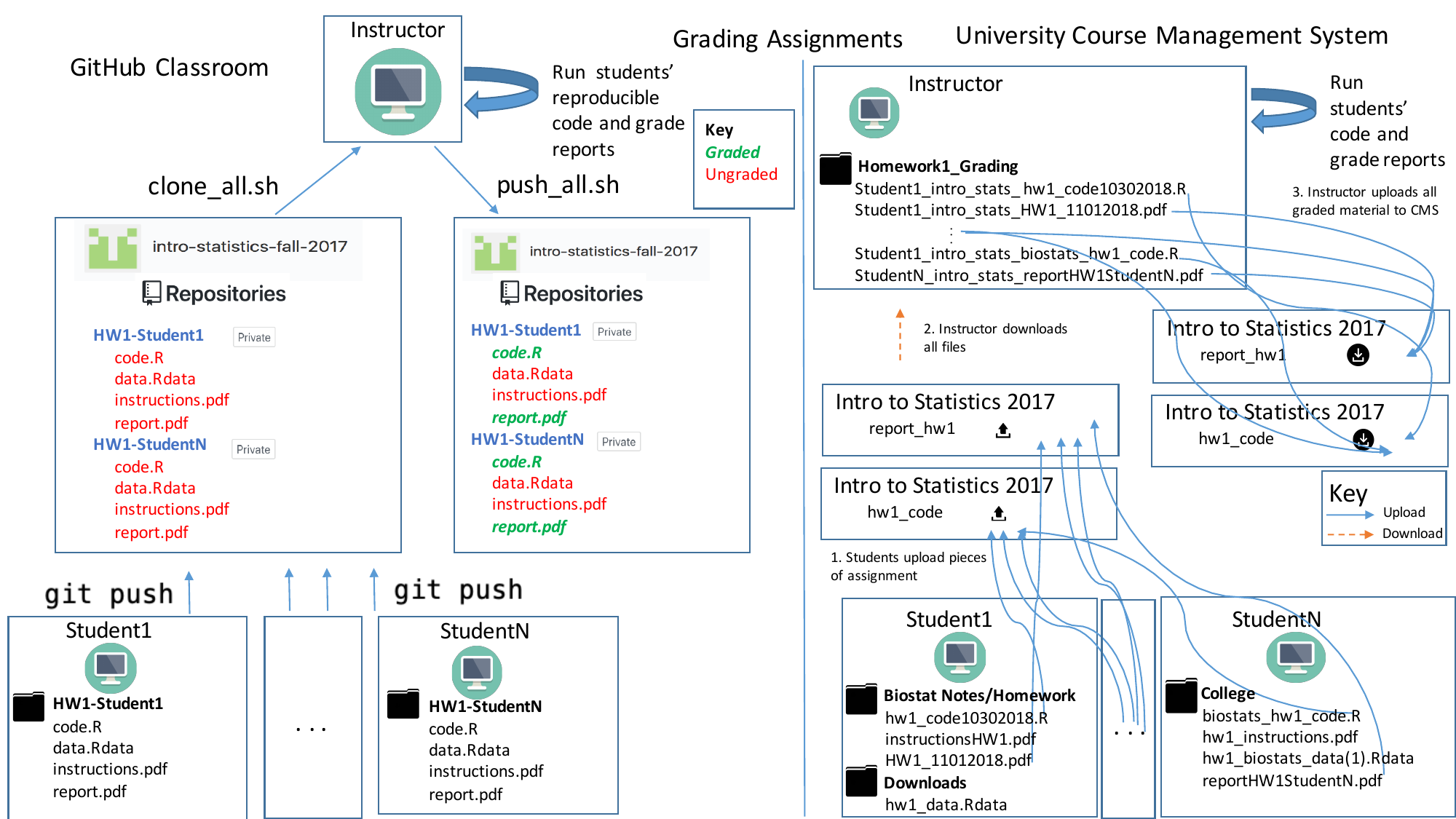}
    \caption{\textbf{Grading assignments with GitHub Classroom vs. a University CMS}. With GitHub Classroom (left side), students all finish the assignment with the same directory structure (bottom left). Students use the {\ttfamily git push} command to upload each piece of their assignment to the GitHub Classroom organization. The instructor then uses our shell script to download all assignments to their local computer, with one command. Because each assignment retains the same directory structure, the instructor can run student code which relies on reading in pieces of data. The instructor then pushes graded assignments with one command back to the GitHub Classroom organization. Using a university CMS, students first have to individually upload each part of the assignment that will be used for grading. Instructors then download each of the uploaded files, and lose all directory structures from students' assignments. After (potentially) running student code and grading, instructors then have to individually upload each graded file back to the CMS.
}
    \label{fig:gh_grade}
\end{figure}

Finally, GitHub Classroom significantly reduces the amount of overhead required to grade and redistribute a large number of assignments (Section \ref{sec:grading}). Figure \ref{fig:gh_grade} visually compares the grading workflow when using GitHub Classroom as compared to an academic CMS. By using GitHub Classroom to collect student assignments, the instructor guarantees that the file structure will be identical in each student's assignment directory, making it easier to check student code for reproducibility. Furthermore, instructors can upload every graded file with one keyboard command, as opposed to individually uploading each graded file back to the CMS for students to access. If there are 8 assignments in a semester, and 50 students, an instructor using their default CMS will have to upload a minimum of 400 individual files, as opposed to an instructor using GitHub classroom who will have to enter only one command per assignment.

In addition to managing assignments through GitHub Classroom, both classes used GitHub Classroom to distribute and update course materials; GitHub can provide course structure (without the teacher relying on their CMS). A student from the ICS lab wrote on their course review, ``I enjoyed the ability to constantly pull updates from the Class Material repository and stay up to date with minimal effort.'' 

Students appreciate the automated tracking of work, for their own knowledge. Throughout the semester, there were many times when students went back to their commit history to see what they had done or to remind themselves about important changes. We cannot overemphasize the importance of tracking student work both for real time knowledge and for practice in using reproducible methods as part of any data analysis.

Finally, we found that students appreciate learning Git and GitHub so that they can apply it to their own research, or so they can become a more attractive job candidate for data science positions. Previous research indicates that students find learning GitHub benefits them in their careers \citep{zagalsky2015emergence}. Although all assignments in both classes were private repositories on GitHub, any work that students want to be made public can be moved to their own public Git presence and can then be shared, e.g., with a prospective employer. Below are some comments on the benefits of GitHub received from students on course reviews for the ICS lab:

\begin{itemize}
\item ``I really enjoyed using GitHub because it's applicable to the other things that I do in my lab''
\item ``I didn't mind including [GitHub] in the curriculum for this semester's course, since it is something that I can now say I know roughly how to use''
\item ``I liked that we can save our past work on GitHub and that it taught me how to take advantage of it for other projects''
\item ``This platform is good for a resum{\'e}́.''	
\end{itemize}

From the ACS course evaluations:

\begin{itemize}
\item ``GitHub is a skill that I think I will value in the future.''
\item ``GitHub was awesome.''
\end{itemize}

\subsection{Motivating the Use of GitHub Classroom}
\label{sec:motivate}
Git and GitHub are often motivated by talking about the importance of version control and reproducibility. While both are indeed huge benefits of using Git and GitHub, they are also abstract concepts to undergraduate students, especially those with minimal statistical and computational experience like the students in the ICS lab. The two courses took different approaches to promoting the benefits of learning Git and GitHub.

In the ICS lab, we motivated the use of Git and GitHub by promoting the advantage of being able to put the skills on a resume. While we also discussed the importance of reproducible research and collaborative coding, the limited research experience of our students meant that ideas of ``best practices'' concepts were more theoretical than practical to them. We emphasized the growing use of tools used for reproducibility in computing and business. 

In the ACS course, we motivated the use of Git through lectures on the reproducibility crisis, giving examples of research projects that went awry \citep{kern2013retraction,chakrabarti2013retraction}. In class, we discuss that Git is not perfect and can be difficult to learn, but that the alternative is more likely to lead to disaster \citep{coombes2007microarrays}. If the students want to pursue quantitative work in fields outside of statistics and data science, such as biology and economics, in today's world, they need to learn how to use today's tools. 

\subsection{Modifying the Introduction of Version Control in Introductory Classes}

To get the most out of introducing GitHub Classroom in an introductory course, we recommend that teachers tailor the extent to which Git and GitHub is used based on the specific goals of the class. While we believe that we still gave students a valuable introduction to Git and GitHub in the ICS lab, in the future we would not introduce these topics at the very beginning of a course targeted towards students without prior computational experience. Instead, we would wait until at least the second half of the course when students have more familiarity with coding and are more comfortable with debugging errors in R (or whichever software is being used in the course). We would have a specific unit on using Git and GitHub for version control, and we would use GitHub Classroom to conveniently distribute assignments to students, giving access to professors and TAs. Students could then focus on learning the Git skills without also having to focus on starting to learn a new coding language. An optimal strategy for integrating the material learned in the Git unit with the rest of the class would be to assign a final project within GitHub Classroom that required students to collaborate in groups using a GitHub repository, demonstrating to students one of the main benefits of GitHub (as discussed in Section \ref{sec:benefits} and in \citet{bryan2018}). 

\section{RESOURCES AND WORKFLOW FOR IMPLEMENTING GITHUB CLASSROOM}
\label{sec:resources}

\subsection{Publicly Available Guides}
\label{sec:guides}
Throughout the start-up and use of GitHub Classroom, we documented our workflows for setting up a GitHub Classroom, sending out assignments, and grading assignments. Our workflows are publicly available in our GitHub Classroom Guide for Teachers \url{https://github.com/jfiksel/github-classroom-for-teachers}. Furthermore, we have made a detailed GitHub Classroom Guide for Students \url{https://github.com/jfiksel/github-classroom-for-students}, complete with written instructions, GIFs, and YouTube videos which will guide students with limited-to-no computing experience through the process of setting up Git and GitHub. In the following sections, we will outline key aspects of our workflow and our guides.

\subsection{Creating a GitHub Classroom}
\label{sec:create}

A GitHub Classroom is organized around GitHub Organizations. A GitHub Organization is a way to share GitHub repositories among multiple users. In creating a classroom, you create an organization for the specific class and semester (i.e. Intro to Statistics Fall 2018), and then link it to the GitHub Classroom software as shown in our guides. Throughout the semester, all student assignments are created as GitHub repositories within the given organization.

Unfortunately, GitHub does not provide a way for students to manually add themselves to the organization, which would reduce the organizational burden of the instructors. To work around this, we recommend using the university's CMS for the course to first send an email to the whole class with a link to an assignment (we will more fully describe this in Section \ref{sec:assignments}). By clicking on the link, students add themselves to the organization.

\subsection{Managing and Creating Assignments}
\label{sec:assignments}

The main purpose of GitHub Classroom is to automate the process of creating and assessing assignments in the form of GitHub repositories for each student (or team). Furthermore, each assignment can be based upon a previously existing repository with starter code.

\begin{figure}[H]
    \centering
    \includegraphics[page=3, width=1.1\linewidth]{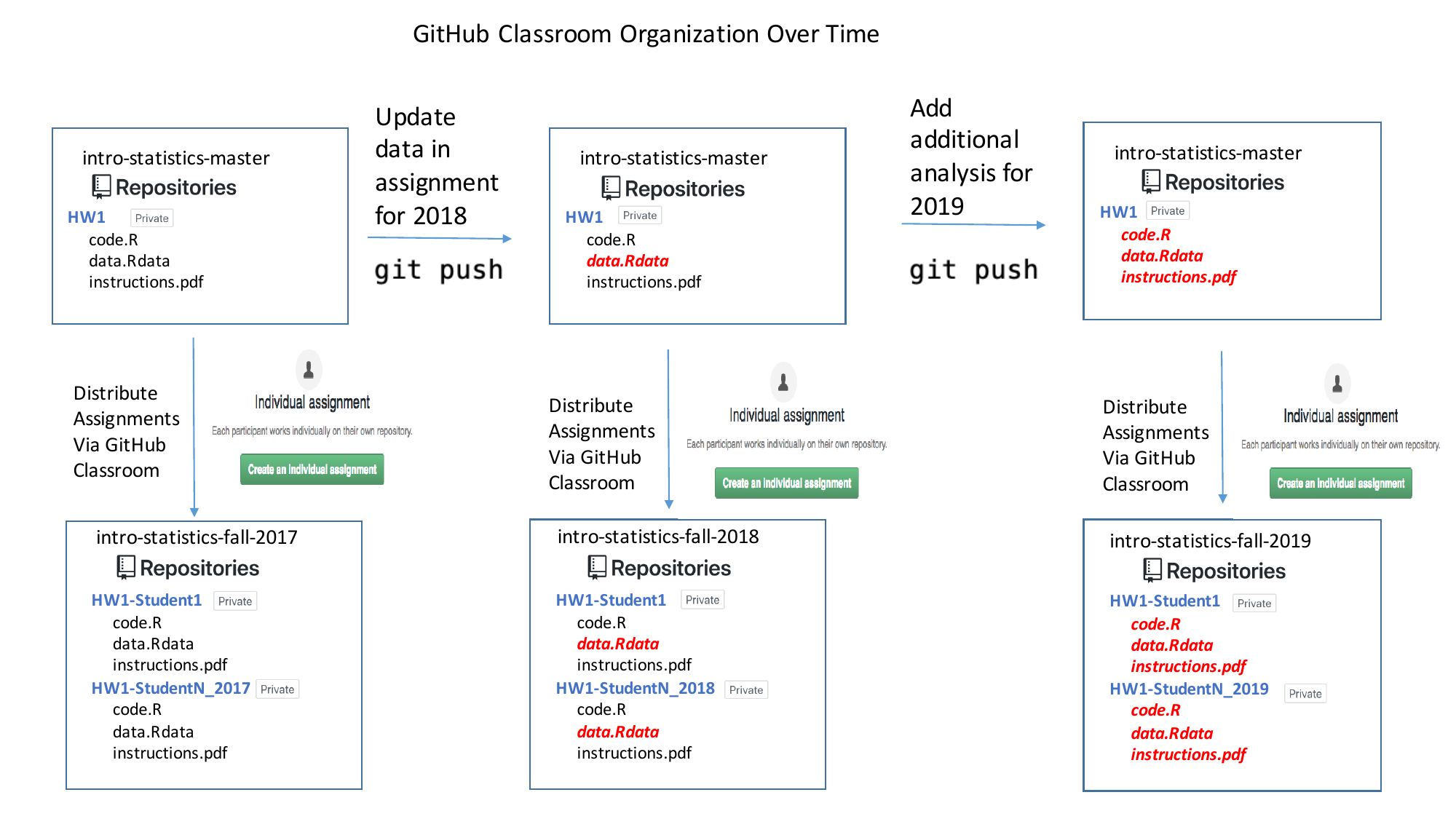}
    \caption{\textbf{Using a master GitHub Classroom organization improves organization of course material}. An instructor who wishes to make a change to an assignment would make a change to the assignment in the master organization (bold and italicized font denotes a changed file), which will then be present when creating new assignments. In this example, the instructor changed the dataset for the 2018 course iteration. In 2019 the instructor added an additional analysis, which resulted in changing the assignment instructions and starter code.
}
    \label{fig:gh_master}
\end{figure}

To best take advantage of this functionality, we recommend having a ``master'' organization (Figure \ref{fig:gh_master}). If you have a class called Introduction to Statistics, you would have one master organization called intro-statistics-master, and then organizations for each iteration of the class (i.e. intro-statistics-Fall-2018, intro-statistics-Fall-2019, etc...). You would then have repositories that contain starter code for each assignment within the master organization. Our guides show how to use starter repositories to reuse the same assignments for each iteration of a class. Another advantage of having a master repository is that you can easily update assignments or revert to old versions of assignments using the Git and GitHub version control tools.

As mentioned in Section \ref{sec:intro}, GitHub allows for unlimited private repositories within education based organizations. GitHub Classroom allows you to make each assignment private with the click of one button. This means that each student (or team) can only view their own repositories, while the instructors have access to all repositories. By using the functionality of GitHub Classroom, instructors not only have an method to (try to) prevent cheating (homework assignments aren't viewed publicly and copied), but can also comply with FERPA privacy rules. If students want to share their work to prospective employers, they can easily make their repositories publicly available. After the final team project in the ACS class (Section \ref{sec:benefits}), several teams made their projects available to the public for increased exposure.

\subsection{Giving Feedback During Assignments}
\label{sec:feedback}

As mentioned in Section \ref{sec:benefits}, one of the main benefits of using GitHub Classroom is the ability to provide feedback in the middle of assignments. If a student is regularly making commits (although this is not a given in an introductory course), instructors can see the specific steps a student has taken to solve a problem. More importantly, however, by cloning a student's GitHub Classroom repository, instructors retain the exact same file structure of data and scripts that the student has. Rather than downloading every piece of data and R script, as described in Section \ref{sec:benefits}, instructors can run the code that the student has implemented to see where the problem with the student's approach lies. The workflow that we employed has the instructor:

\begin{enumerate}
\item Clone the directory to their own computer. Or pull the latest changes if the repository has been previously cloned.
\item Provide feedback through either a pull request, directly into the code, or as an issue in GitHub.
\item Push all changes back to GitHub.
\end{enumerate}

All feedback is then documented in the commit history, which can be a useful reference if the student wants to look back at the assignment after completion. 

\subsection{Grading Workflows}
\label{sec:grading}

We have identified two potential ways to grade assignments from GitHub Classroom. The first method, which we did not use, is through the use of Pull requests directly on GitHub. This method is detailed in a GitHub blog post \url{https://blog.github.com/2017-06-13-how-to-grade-programming-assignments-on-github/}. The advantage to this approach is that instructors can provide line-by-line code comments, along with syntax highlighting, that is easily viewable by the student after grading. The immediate downside is that it requires a decent amount of manual overhead work, as the instructor has to click through each assignment and go through the pull request process on GitHub. GitHub Organizations do not organize assignment repositories in a convenient way (yet), and the number of repositories to keep track of can quickly become overwhelming -- in the end, the ICS lab had 256 repositories in their class organization, and the ACS course had 878 repositories. GitHub classroom makes {\it tracking} repositories more manageable by allowing you to click on an assignment inside of the GitHub Classroom portal and then list all student repositories for this assignment, but it does not have an automated way of {\it pulling} the large number of repositories. Yet another downside to the ``feed back through pull requests" approach is that the instructor does not have the code locally to run. Integration tools such as Werker \citep{rundel2018} can automatically check for reproducibility of a student's assignment directly from GitHub, meaning instructors do not have to clone assignments onto their local computers to run the code.

The second method for grading assignments, which we took advantage of, used a shell script \url{https://github.com/jfiksel/mass_clone}  (modified from \url{https://github.com/konzy/mass_clone}) to automatically clone all assignments to our local computers. We could then run all student code locally and provide comments directly on each student's coding script. Detailed instructions, including optimal directory set-up, is available under the ``Grading assignments" section of our GitHub Classroom Guide for Teachers. We give our method step-by-step below: 
\begin{enumerate}
\item{Clone all student assignment repositories to a local computer using our shell script}
\item{Open each student's assignment (which can be done with a single command on the command-line), and run the code inside of RStudio to ensure reproducibility. Add comments, suggestions, or edits within each student's .R script or .Rmd file and save the altered file. }
\item{Use the shell script in step 1 to simultaneously add, commit, and push all edits for all student assignments. This step is done with one line of "command line" code, and students can view highlighted comments by visiting their assignment repository on GitHub and clicking on the commit, or by pulling the latest version of the repository.}
\end{enumerate}

The choice of grading method will most likely be determined by course size, emphasis on code reproducibility, and individual preference. For courses where grading is assisted or done solely by teaching assistants, instructors will have to ensure that teaching assistants are comfortable with GitHub, Git, and have basic understanding of the command line. 

\subsection{Using GitHub Classroom to Distribute Lecture Materials}

Both classes used GitHub Classroom not only for assignment distribution and grading but also for updating and sharing course material. Since class periods often consisted of live coding, lecture material was updated in real-time and could not be shared in its final version until after each class session was complete. So, in addition to the assignment repositories, we also each maintained a single repository within the class organization for class materials. Our guides describe how to give all students access to the class materials repository while also keeping it private from GitHub users outside of the class. In our set-up, the class materials repository has a folder for each class meeting, containing starter code and data to be used that day. Students pulled this material before class to follow along. After updating the materials with code written during the class meeting, the instructor would push the new material to the repository, and students could again pull the material to get the complete files produced during the class. This not only helped students stay up to date with the class, but it also gave them weekly practice in using GitHub. 

\section{CONCLUSION}
\label{sec:conc}
Previous work has shown that GitHub can be used for educational purposes across a range of subjects, class sizes, and instructor knowledge of GitHub \citep{zagalsky2015emergence}. The contribution of our work is to provide a concrete and easy-to-implement workflow for instructors who want to bring version control into their classroom. By using our recommended workflow, instructors will not only benefit their students by teaching them skills desired by potential employers but will also significantly cut down on the administrative work required to distribute, grade, and return assignments. By saving time that was formerly spent on administrative duties, instructors can spend more time working with students and updating course material. 

Through our experiences, we show that the Git workflow can be used in both introductory and advanced courses and by instructors without previous GitHub experience. We have used student feedback to construct a separate guide to Git and GitHub for students, so that instructors unfamiliar with version control do not have to create their own teaching materials. Our paper also discusses potential ways of introducing and motivating version control to students, which adds to the discussion in \citet{zagalsky2015emergence} about various motivations instructors have for using GitHub for education. Our hope is that our guides serve as a starting point for instructors to use GitHub, who will then modify and improve our workflows for different class settings.

\bibliographystyle{chicago}
\bibliography{github-classroom}
\end{document}